\documentstyle[12pt]{article}
\begin{document}

\small
\hoffset= -1truecm
\voffset= -2truecm
\title{\bf The nontopological scalar solitons in de Sitter spacetimes}
\author{Hongbo Cheng\footnote{E-mail: hbcheng@public4.sta.net.cn}\hspace{0.6cm} Zhengyan Gu\\\footnotesize \it
        Department of Physics, East China University of Science and Technology, Shanghai 200237, P. R. China}
\date{}
\maketitle

\begin{abstract}
  In this letter the nontopological scalar solitons with a self-interaction potential are investigated in a
  static de Sitter spacetime and their field
  equations are derived. In particular, with series expansion we prove analytically the nonexistence of the
  solutions under the boundary conditions, which show that some
  nontopological solitons can not live in the background.
\end{abstract}
\vspace{8cm} \hspace{1cm}
\newpage

\noindent \textbf{1.}\hspace{0.4cm}\textbf{Introduction}

The recent astronomical observations indicated that the
cosmological constant in our universe is positive [1, 2]. The de
Sitter spacetimes have been receiving more attention [3-7]. As an
anomaly, the de Sitter spacetime is inconsistent with
supersymmetry and can not be embedded into string theory. In
addition, many problems such as event horizon, entropy etc. are
still puzzling. We have to undergo the extensive studying de
Sitter spacetimes in various directions.

Researching nontopological solitons (NTS) with self-interaction
potential is valuable and also become a focus. Many stable NTS
including Q-balls etc. arise in various field theories [8, 9] and
are also cosmologically significant [10-13]. They are natural
candidates for dark matter and may be an important factor in
considering the neutron stars, etc., more applications to
astrophysics.

A feasible and effective method was adopted in order to study
further the de Sitter space. Some topological defects have been
investigated in the backgrounds [14,15]. The models such as global
topological defects and local vortices satisfying the conditions
can exist and their solution shapes depend on the characteristics
of de Sitter spacetimes. The structures of spacetimes have great
influence upon them.

The relationship between NTS and de Sitter space is very
significant and must be shown. A lot of achievements in
researching NTS for cosmology has been made, which show that the
solitons will become the promising candidates for dark matter. It
is necessary to explore NTS in de Sitter spacetimes with positive
cosmological constant supported by the observational evidence. We
have to derive the equations of motion of NTS in de Sitter
spacetime and discuss the solutions within the horizon. An
effective and reliable method is chosen to find that NTS with the
self-interaction potential can not exist but only the trivial
solution can although the NTS are more important to astrophysics.
The result is completely different from topological defects in the
same surroundings.

In this letter, we derive the field equations for NTS at first and
solve that under the stationary ansatz and the necessary boundary
conditions with series expasion. The solutions in series form are
obtained and discussed in the region $r\in[0,L]$, where $L$ is the
de Sitter radius. The conclusions are emphasized at last.

\vspace{0.8cm}
\noindent\textbf{2.}\hspace{0.4cm}\textbf{Nontopological scalar
solitons in de Sitter spacetimes}

We start to consider the NTS in the background of de Sitter (dS)
spacetime. The metric of four-dimensional static dS spacetime is

\begin{equation}
ds^{2}=(1-\frac{r^{2}}{L^{2}})dt^{2}-\frac{1}{1-\frac{r^{2}}{L^{2}}}dr^{2}-r^{2}(d\theta^{2}+\sin^{2}\theta
d\varphi^{2})
\end{equation}

\noindent where the cosmological constant
$\Lambda=\frac{3}{L^{2}}$, $L$ the de Sitter radius. Within the
horizon the coordinate $r$ is allowed to be $0\leq r\leq L$. Here
the Lagrangian for NTS can be chosen as

\begin{equation}
{\cal L}=\sqrt{-g}(\partial_{\mu}\Phi\partial^{\mu}\Phi -U(\Phi))
\end{equation}

\noindent in general, the potential
$U(\Phi)=m^{2}\Phi\Phi^{\ast}+\sigma(\Phi\Phi^{\ast})^{2}+\lambda(\Phi\Phi^{\ast})^{3}$
with a global minimum $U(0)=0$ at $\Phi=0$. $m$ is the "bare" mass
of the boson and $\sigma$, $\lambda$ are coupling constants. The
stationary ansatz describing the spherically symmetric bound state
of scalar field with frequency $\omega$ is

\begin{equation}
\Phi(t,\mathbf{r})=P(r)e^{-i\omega t}
\end{equation}

\noindent By using the ansatz (3), the field equation can be
reduced to,

\begin{equation}
(1-x^{2})\frac{d^{2}P}{dx^{2}}+2(\frac{1}{x}-2x)\frac{dP}{dx}+\frac{\omega^{2}L^{2}}{1-x^{2}}P-m^{2}L^{2}P-2\sigma
L^{2}P^{3}-3\lambda L^{2}P^{5}=0
\end{equation}

\noindent where

\begin{equation}
x=\frac{r}{L}\in[0,1]
\end{equation}

\noindent The solution must obey the boundary conditions,

\begin{equation}
P(x=0)=P_{0}, \hspace{2cm} P(x=1)=0
\end{equation}

\noindent where $P_{0}$ is an arbitrary constant. Solving equation
(4) explicitly by the series in the form

\begin{equation}
P(x)=\sum_{n=0}^{\infty} a_{2n}x^{2n}
\end{equation}

\noindent In the region $x\in[0,1)$, the recursion formula of the
coefficients $a_{k}$ can be expressed as follow

\begin{equation}
a_{0}=P_{0}
\end{equation}

\begin{equation}
a_{2}=\frac{a_{0}L^{2}}{6}\bigg[3\lambda a_{0}^{4}+2\sigma
a_{0}^{2}+(m^{2}-\omega^{2})\bigg]
\end{equation}

\begin{eqnarray}
2(n+2)(2n+5)a_{2n+4}&=&[(8(n+1)(n+2)+(m^{2}-\omega^{2})L^{2})]a_{2n+2}\nonumber\\
& &-[2n(2n+3)+m^{2}L^{2}]a_{2n}\nonumber\\
& &+2\sigma L^{2}\sum_{j+k+l=2n+2} a_{j}a_{k}a_{l}-2\sigma
L^{2}\sum_{j+k+l=2n} a_{j}a_{k}a_{l}\nonumber\\
& &+3\lambda L^{2}\sum_{j+k+l+p+q=2n+2}
a_{j}a_{k}a_{l}a_{p}a_{q}\nonumber\\
& &-3\lambda L^{2}\sum_{j+k+l+p+q=2n} a_{j}a_{k}a_{l}a_{p}a_{q}
\end{eqnarray}

\noindent where $j$, $k$, $l$, $p$, $q=0,2,4, \cdot\cdot\cdot$,
$n=0,1,2,\cdot\cdot\cdot$. On the other hand, $0\leq x\leq 1$,
near $x=1$ asymptotic region, the solution to equation (4) becomes
in the form of series,

\begin{equation}
P(x)=\sum_{n=0}^{\infty} b_{n}(1-x)^{n}
\end{equation}

\noindent According to the boundary conditions (6), $b_{0}=0$. We
show that the recursion formula of the coefficients $b_{n}$ can be
expressed as follow,

\begin{equation}
b_{1}=0
\end{equation}

\begin{equation}
b_{2}=0
\end{equation}

\begin{eqnarray}
[4(n+3)^{2}+\omega^{2}L^{2}]b_{n+3}&=&[2(n+2)(13n+22)+(\omega^{2}+2m^{2})L^{2}]b_{n+2} \nonumber\\
& &-[(n+1)(5n+16)+3m^{2}L^{2}]b_{n+1} \nonumber\\
& &+[n(n+3)+m^{2}L^{2}]b_{n} \nonumber\\
& &+2\sigma^{2} L^{2}\sum_{j+k+l=n}
b_{j}b_{k}b_{l}-6\sigma^{2}L^{2}\sum_{j+k+l=n+1}
b_{j}b_{k}b_{l}\nonumber\\
& &+4\sigma^{2}L^{2}\sum_{j+k+l=n+2} b_{j}b_{k}b_{l}+3\lambda
L^{2}\sum_{j+k+l+p+q=n}
b_{j}b_{k}b_{l}b_{p}b_{q}\nonumber\\
& &-9\lambda L^{2}\sum_{j+k+l+p+q=n+1}
b_{j}b_{k}b_{l}b_{p}b_{q}\nonumber\\
& &+6\lambda L^{2}\sum_{j+k+l+p+q=n+2} b_{j}b_{k}b_{l}b_{p}b_{q}
\end{eqnarray}

\noindent where $j$, $k$, $l$, $p$, $q$,
$n=0,1,2,\cdot\cdot\cdot$. The consequence is $b_{n}=0$,
$n=0,1,2,\cdot\cdot\cdot$. From expression (11), we obtain

\begin{equation}
P(x\leq 1)=0
\end{equation}

The trivial solution shows that fields are in the vacuum states in
the region $x\in(0,1]$, so the NTS can not appear here either.
According to equations (7~10) and (15), there is no smooth and
nontrivial solution to the field equation (4) under the boundary
conditions (6) at $x\in [0,1]$, which reveals that the solutions
such as nontopological solitons can not exist in the de Sitter
spacetimes.

\vspace{0.8cm} \noindent
\textbf{3.}\hspace{0.4cm}\textbf{Conclusion}

Here we prove the nonexistence of some types of solutions such as
nontopological scalar solitons with a self-interaction potential
in a static de Sitter background in rigorous manner. Although many
scalar field equations in the presence of a horizon were
considered [16], but many kinds of equations need to be studied in
detail respectively under different environments if it is
necessary. It is essential for us to derive the field equation
with continuous spectrum and solve that with larruping method of
series expansion.Our expressions show convincingly that only
trivial solution can exist, which prove analytically that no NTS
can inhabit in the de Sitter spacetime in contrast to the
topological defects [14, 15]. The interesting results encourage us
to continue studying NTS in other different spacetimes.

\vspace{0.8cm}
\noindent \textbf{Acknowledgement}

This work was supported by the Basic Theory Research Fund of East
China University of Science and Technology, grant No. YK0127312.

\end{document}